\documentclass[11pt,a4paper]{article}
\usepackage[T1]{fontenc}     
\usepackage[utf8]{inputenc}  % Accents codés dans la fonte
\usepackage[french, english]{babel}  % Les traductions françaises
\usepackage{numprint}        % \numprint(9,36) pour utilisation de la virgule comme séparateur décimal

\usepackage{amsmath}         % Les maths de base
\usepackage{esint}
\usepackage{amssymb}
\usepackage{outlines}
\usepackage{cite}

\everymath{\displaystyle}

\usepackage[svgnames]{xcolor}% Pour les besoins de PythonTeX
\usepackage[margin = 2cm]{geometry}        % Gestion des dimensions des pages
\usepackage{soul}
\usepackage{stmaryrd}

\usepackage{setspace}

\usepackage{pythontex}       % Utilisation de PythonTeX

\usepackage{graphicx}        % Gestion des inclusions graphiques
\usepackage{comment}
\usepackage{float}
\usepackage{subcaption}

\usepackage{tensor}

\usepackage{listings}
\usepackage{hyperref}

\usepackage{tikz}            % Si on veut présenter le code Python
\usepackage[framemethod=TikZ]{mdframed}
\newcommand\cA{h}
\newcommand\cB{f}
\newcommand\h{h}
\newcommand\f{f}

\newcommand\Qa{\hat Q}

\title{Slowly rotating Black Holes in DHOST Theories}

\author{Hugo Candan$^{a,b}$ , Karim Noui$^{b}$ and David Langlois$^{c}$  \\ \\
\small $^{a}${\it{LUX, Observatoire de Paris, Universit\'e PSL, Sorbonne Universit\'e, CNRS, 92190 Meudon, France}}\\
\small $^{b}${\it{Universit\'e Paris-Saclay, CNRS/IN2P3, IJCLab, 91405 Orsay, France}}\\
\small $^{c}${\it{Université Paris Cit\'e, CNRS, Astroparticule et Cosmologie, F-75013 Paris, France}}
}
\date\today

\begin{document}
\maketitle

\begin{abstract}
    We study slowly rotating black hole solutions within Degenerate Higher Order Scalar Tensor (DHOST) theories. Starting from a static, spherically symmetric metric solution of a DHOST theory, we employ the Hartle–Thorne ansatz to model a slowly rotating spacetime. We show that the differential equation governing the  frame-dragging function $\omega$ (which is supposed to depend on the radial coordinate only) is integrable for any DHOST theory allowing us to obtain its explicit form. We also consider angular dependence in $\omega$ and show that, for specific DHOST theories, regularity at the horizon and at infinity forbids it, as in General Relativity. As an illustration of the formalism introduced here, we study the slowly-rotating version of  black hole solutions with primary hair obtained recently, examining the influence of the rotation on the Innermost Stable Circular Orbit (ISCO) and on the 
    circular light trajectories in the equatorial plane.
\end{abstract}
%\tableofcontents

\setcounter{figure}{0}

\normalsize

\singlespacing

\section{Introduction}

\hspace{\parindent} 

General Relativity (with a cosmological constant) is the unique theory that satisfies the assumptions of Lovelock’s theorem in four dimensions \cite{Lovelock:1971yv}. However, there are many reasons to consider theories beyond General Relativity, whether phenomenological or more theoretical. One of the simplest ways to construct alternatives to Einstein’s gravity is to assume that the metric field comes with a scalar field to mediate the gravitational interaction. Scalar–tensor theories have been extensively studied in recent years, leading to the classification of Degenerate Higher Order Scalar-Tensor (DHOST) theories \cite{Langlois:2015cwa,Langlois:2015skt,BenAchour:2016fzp} (and even more generally U-DHOST theories which are degenerate in the unitary gauge \cite{DeFelice:2018ewo}), which currently represent the most general class of theories propagating only three degrees of freedom. In physically relevant cases, these correspond to the two tensorial degrees of freedom of the metric field and one additional scalar degree of freedom.
DHOST theories have been %extensively studied and 
applied to cosmological and astrophysical   contexts (see the review \cite{Langlois:2018dxi}), such as  black holes or compact objects.

In General Relativity (GR), black holes are known to have "no hair" \cite{Carter:1971zc}. The only stationary and axisymmetric solution of the vacuum Einstein equations is the Kerr metric \cite{Kerr:1963ud}, described by two parameters, its mass $M$ and its angular momentum $J$. In the context of scalar-tensor theories, no hair theorems have also been derived for some subclasses of DHOST theories, see for instance \cite{Hui:2012qt} and \cite{Capuano:2023yyh}, but their assumptions are generally more restrictive compared to GR. In the recent years, many hairy static black hole solutions have been found within DHOST theories, breaking the hypotheses of these theorems, as 
illustrated
%it can be seen 
in \cite{Babichev:2017guv,Babichev:2023psy, Bakopoulos:2023fmv} for instance. These solutions exhibit a non-trivial scalar hair, and the spacetime metric can differ from the Schwarzschild solution expected in GR. However, very few solutions describing a rotating black hole are known to date. One of the first rotating solutions are the so-called "stealth Kerr" solutions which were found in \cite{Babichev:2017guv,Charmousis:2019vnf} by assuming the spacetime metric to be that of Kerr, and then solving for the scalar field. 
Starting from a stealth Kerr solution, one can apply a disformal transformation to the metric to obtain a new, non-stealth, solution, as shown in \cite{Anson:2020trg, BenAchour:2020fgy, Achour:2021pla, Babichev:2024hjf,BenAchour:2025lkx}. Among its notable features, the resulting metric is generically non-circular \cite{Babichev:2025szb}. These solutions currently constitute the only fully analytic examples of rotating black holes with scalar hair in DHOST theories. By contrast, non–disformally related Kerr solutions have been obtained only numerically, for instance in the cubic Galileon theory \cite{VanAelst:2019kku, Grandclement:2023xwq} and in scalar–Gauss–Bonnet theories \cite{Kleihaus:2011tg, Kleihaus:2015aje, Delgado:2020rev}.

The limited number of analytical rotating solutions stems from the fact that solving the equations of motion becomes considerably more challenging when the spherical symmetry assumption of the static case is abandoned. However, analytical approximate solutions can still be obtained and approximate rotating black hole solutions in some specific scalar-tensor theories have already been found. An analytical slowly rotating black hole solution in a scalar–Gauss–Bonnet theory was presented in \cite{Charmousis:2021npl} and further analyzed in \cite{Gammon:2022bfu}. Moreover, for a specific subclass of DHOST theories, a no-hair theorem at first order in the angular momentum 
$J$ has been derived in \cite{Maselli:2015yva}.

In this paper, we present general solutions for slowly rotating black holes in DHOST theories. Specifically, we make use of the well-known Hartle–Thorne ansatz \cite{Lense:1918, Hartle:1967he, Hartle:1968si}, originally introduced by Lense\footnote{Thanks to David Kubz\v{n}\'ak for giving us the reference to the article of Lense and Thirring} and subsequently by Hartle and Thorne to describe the gravitational field of a rotating relativistic star at first order in the angular momentum $J$
within the framework of General Relativity. We show that, at first order in $J$, the equations of motion exhibit a remarkable integrability property, which allows us to derive a very general expression for slowly rotating black hole metrics in DHOST theories in terms of the corresponding static solutions. Notably, this integrability property is likely not exclusive to scalar–tensor theories, and the method presented here could potentially be extended to other modified gravity frameworks. When specializing this general result to an interesting subclass of theories possessing shift and parity symmetry, we find  that the first-order rotational term is always identical to that of Kerr, even when the static solution differs from Schwarzschild. In order to illustrate these results,
we study the slowly-rotating version of black hole solutions with primary hair obtained recently in \cite{Charmousis:2025xug}, and we 
examine the influence of the rotation on the Innermost Stable Circular Orbit (ISCO) and on the circular light trajectories in the equatorial plane.

Let us emphasize that the study of rotating black holes in modified gravity theories is particularly interesting as we are collecting a lot of data from compact objects in our universe, using both gravitational wave detectors from the LIGO-Virgo-KAGRA collaboration \cite{LIGOScientific:2016aoc,LIGOScientific:2017zic}, and direct imaging by the Event Horizon Telescope \cite{EventHorizonTelescope:2019dse}. Up to this day, all the data collected from these compact objects are compatible with them being black holes, or neutron stars, in General Relativity \cite{LIGOScientific:2025obp,LIGOScientific:2025rid}, but 
 possible deviations from GR will be further explored  with
more precise measurements, for instance with future detectors LISA \cite{Barausse:2020rsu} or Einstein Telescope \cite{ET:2025xjr}.

The paper is organized as follows. The next section is devoted to the derivation of slowly rotating solutions in DHOST theories. We simplify the equation governing the frame-dragging function $\omega$, present a general integral solution, and discuss the origin of the integrability property of this equation. 
In Section~3, we apply our results to shift-symmetric theories, for which the expression of $\omega$ simplifies. We also study how slowly rotating solutions transform under disformal transformations and examine the impact of rotation on the innermost stable circular orbit (ISCO) and on the circular light trajectories in the equatorial plane. 
In Section~4, we investigate a possible angular dependence of $\omega$ and show that regularity conditions at the horizon and at infinity forbid any $\theta$-dependence of $\omega$. We conclude with a brief summary and provide appendices containing technical details.

\section{Slowly rotating solutions in DHOST theories}
\hspace{\parindent} After a brief review of DHOST theories, we study their solutions for slowly rotating space-time following the Hartle-Thorne ansatz. Interestingly, we find  an explicit integral formula for the first order term in the slowly rotating metric, for any (quadratic and cubic) DHOST theory.

\subsection{DHOST theories: brief review and notations}
\hspace{\parindent} DHOST theories,
introduced and classified in \cite{Langlois:2015cwa,Langlois:2015skt,BenAchour:2016fzp} (see also \cite{Langlois:2018dxi,Kobayashi:2019hrl} for reviews),  are scalar-tensor theories  that  propagate only three degrees of freedom (two tensor modes and one scalar mode) even though their Lagrangian can involve non-trivial second-order derivatives of the scalar field\footnote{DHOST theories can be complemented by  U-DHOST theories \cite{DeFelice:2018ewo} which are degenerate in the unitary gauge where the scalar field is used to define the time coordinate. }. The absence of a  pathological extra scalar mode  is ensured by the degeneracy conditions imposed on the functions appearing in the Lagrangian.

The  DHOST  action, up to  cubic terms in the second derivatives of the scalar field $\phi$, takes the form
\begin{equation}
\begin{split}
\label{DHOSTaction}
    S[g_{\mu\nu},\phi]=\int \mathrm{d}^4x ~\sqrt{-g}~ \Bigg[&F_0(\phi,X)+F_1(\phi,X)\Box\phi+ F_2(\phi,X)R+ F_3(\phi,X)G^{\mu\nu}\nabla_\mu\nabla_\nu\phi\\&+\sum_{i=1}^5A_i(\phi,X) \, {L}^{(2)}_i+\sum_{i=1}^{10}B_i(\phi,X) \, {L}^{(3)}_i\Bigg] \, ,
\end{split}
\end{equation}
where we are using the convention\footnote{The alternative convention  $X=\partial_\mu\phi\partial^\mu\phi$ is also often encountered in the literature, in particular in \cite{Langlois:2015cwa,Langlois:2015skt,BenAchour:2016fzp}. } $X =-({1}/{2})\partial_\mu\phi\partial^\mu\phi$ for the canonical kinetic term of the scalar field. 
The expressions for the elementary quadratic and cubic Lagrangians ${L}_i^{(2)}$ and ${L}_i^{(3)}$ can be found in \cite{BenAchour:2016fzp} for example. The action is parametrized by the functions $F_i$, $A_i$ and $B_i$, which are in general  functions of $\phi$ and $X$ and  must  satisfy  degeneracy conditions, given explicitly in \cite{Langlois:2015cwa,BenAchour:2016fzp}.

\subsection{The Hartle-Thorne ansatz for the metric}
\hspace{\parindent} 
Many static and spherically symmetric solutions (black hole and more generally exotic compact objects) have been discovered (analytically and numerically) in the context of DHOST theories \cite{Kanti:1995vq,Rinaldi:2012vy,Babichev:2013cya,Sotiriou:2014pfa,Babichev:2017guv,Lu:2020iav,Bergliaffa:2021diw,Bakopoulos:2023fmv}. By contrast, very few rotating (axisymmetric) solutions have been reported in the literature. The first rotating solution found analytically was the stealth Kerr solution\cite{Babichev:2017guv,Bakopoulos:2023fmv} where the metric is identical to the Kerr metric. From the stealth Kerr solution, new solutions were constructed \cite{BenAchour:2020fgy,Anson:2020trg} by applying a disformal transformation (see section \ref{disformal}). Numerical solutions have also been found, for the cubic galileon theory \cite{VanAelst:2019kku,Grandclement:2023xwq}, and scalar-Gauss-Bonnet theories \cite{Kleihaus:2011tg,Kleihaus:2015aje,Delgado:2020rev}. 

Here, we study solutions that describe a rotating black hole space-time with a non-trivial scalar field profile at  first order in the angular momentum $J$. Starting from an arbitrary static and spherically symmetric metric solution 
$g^{(0)}_{\mu\nu}$,   we look 
%then  
for a new solution of the form
\begin{equation}
g_{\mu\nu} = g^{(0)}_{\mu\nu}+J \, g^{(1)}_{\mu\nu}+{\cal O}(J^2)\,.
\end{equation}
Following the Hartle-Thorne ansatz, we assume the new metric to be of the form
\begin{equation}
\label{hartlethornemetric}
    \mathrm{d}s^2=-\h(r)\mathrm{d}t^2+\frac{\mathrm{d}r^2}{\f(r)}+r^2(\mathrm{d}\theta^2+\sin^2(\theta)\mathrm{d}\varphi^2)-2\omega(r)r^2\sin^2(\theta)\mathrm{d}t\mathrm{d}\varphi +{\cal O}(J^2)\,,
\end{equation}
where the function $\omega(r)$ is implicitly  proportional to  $J$, and characterizes the frame dragging effect.

To our knowledge, this ansatz was first introduced by Lense and Thirring in 1918 \cite{Lense:1918}, and was popularized later by Hartle and Thorne to describe the gravitational field of a rotating star within General Relativity \cite{Hartle:1967he,Hartle:1968si}. One may wonder how general this ansatz is, in particular whether any stationary rotating spacetime reduces to this form at linear order in the angular momentum. The expression \eqref{hartlethornemetric} was derived by Hartle using the specific symmetry of circular spacetimes \cite{Hartle:1967}, but it actually also works for the non-circular solution found in \cite{BenAchour:2020fgy,Anson:2020trg}, suggesting that it might be more general. The possible generalization by adding a dependence of the function $\omega$ on the angle $\theta$ will be discussed in section \ref{angulardep}.

At this stage, we follow the usual procedure, which consists in, first, substituting the ansatz \eqref{hartlethornemetric} into the equations of motion for the metric, namely
\begin{eqnarray}
  \mathcal{E}_{\mu\nu} \equiv \frac{1}{\sqrt{-g}}\frac{\delta S}{\delta g^{\mu\nu}}  = 0 \, ,
\end{eqnarray} and then in expanding these equations up to linear order in $\omega$. Notice that we do not need to take into account the equation of motion for the scalar field since it is a consequence of the metric equations of motion.

In theories invariant under the global shift symmetry $\phi\to \phi+$const, which will be discussed in section \ref{shiftsymmetry}, the scalar field can acquire a linear time dependence, 
\begin{equation}
\label{phiexpression}
    \phi = qt+\psi(r)\,,
\end{equation}
without spoiling the stationarity of the metric, because all physical quantities in this case depend on the gradient of $\phi$ and not $\phi$ itself (see for instance \cite{Babichev:2013cya}). In the most general case, where the theory is not shift-symmetric, the parameter $q$ must vanish. Combining \eqref{phiexpression} with the metric ansatz \eqref{hartlethornemetric}, and expanding at first order in $\omega$ gives the following expression for $X$,
\begin{equation}
    X =\frac{q^2}{2h}-\frac{1}{2}f\psi'^2\,,
\end{equation}
which happens to be independent of $\omega$.

Let us recall that, in the absence of rotation (when $\omega=0$), the diagonal equations  $\mathcal{E}_{tt}$, $\mathcal{E}_{rr}$ together with the non-diagonal one $\mathcal{E}_{tr}$ (or equivalently the equation of motion of the scalar field) enable us to determine the two metric components $\h(r)$ and $\f(r)$ and the scalar field $\psi(r)$. All the other equations can be shown to be redundant.

In the presence of a slow rotation, one can easily see that $\mathcal{E}_{tt}$, $\mathcal{E}_{rr}$ 
and $\mathcal{E}_{tr}$
remain unchanged compared to the case where $\omega=0$ (they do not involve terms linear in $\omega$), and are thus satisfied by  the  metric components $\h(r)$ and $\f(r)$ of the static solution. Furthermore, as scalar quantities can only depend on even powers of the angular momentum $J$, the scalar field $\phi$ also remains unchanged in the slow rotation approximation.

The difference with respect  to the static case arises from the non-diagonal equation $\mathcal{E}_{t\varphi}$, which is no longer trivial and yields  a second order linear differential equation for $\omega(r)$, with coefficients depending on the static solution, i.e. on the functions $\h(r)$, $\f(r)$ and $\psi(r)$. Hence $\omega(r)$ can be obtained by solving directly $\mathcal{E}_{t\varphi}=0$, but we will see that it is more convenient to consider a linear combination of $\mathcal{E}_{t\varphi}$ and $\mathcal{E}_{\varphi\varphi}$ which is straightforward to integrate.

Before discussing DHOST theories, let us recall that, for  the Kerr  metric in General Relativity, the function $\omega$ is given by 
\begin{equation}
\label{omegakerr}
    \omega_{\mathrm{Kerr}}(r) = \frac{2J}{r^3} \,,
\end{equation}
where $J$ is the Arnowitt–Deser–Misner (ADM) angular momentum of the black hole\footnote{The expression \eqref{omegakerr} generalizes to $$\omega(r) = \frac{2J}{r^3} +\frac{J\Lambda}{3}$$ for Kerr-(A)dS slowly rotating metric, where $\Lambda$ is the cosmological constant}. 

In the following, the integration constant appearing in $\omega(r)$ when solving the equations will be expressed in term of $J$  to match the asymptotic behavior of the Kerr solution, whenever possible.

\subsection{Solutions for DHOST theories}
    \hspace{\parindent} 
    We now discuss the specific case of DHOST theories, which includes most known scalar-tensor theories.
    To write and solve the equation solved by the function $\omega(r)$, it is convenient to consider the linear combination  $\mathcal{E}_{t\varphi}+\omega\mathcal{E}_{\varphi\varphi}$, which can be shown to reduce to the very simple form
    \begin{equation}
    \label{eom}
        \mathcal{E}_{t\varphi}+\omega\mathcal{E}_{\varphi\varphi} = \frac{1}{2r^2}\sqrt{\h\f}\sin^2(\theta)\frac{\mathrm{d}}{\mathrm{d}r}\bigg[Q(r)\omega'(r)\bigg]\,,
    \end{equation}
    with $Q(r)$    given by 
    \footnote{The expression of $Q$, 
    %can be extended to higher order scalar tensor theories which are not necessarily degenerate, in which case
    starting from the action \eqref{DHOSTaction}, is given by
    \begin{equation}
    \label{Qnondegen}
    \begin{split}
        Q = &\sqrt{\frac{\f}{\h}}\bigg[(F_2+2XA_1)r^4+\f\,\psi'X(4B_2+3B_3)r^3\\&-\frac{1}{2}\f\,\psi'X'(F_{3X}+3B_3+4XB_6)r^4-\frac{X}{\h\psi'}(X\h)'(3B_3+2B_2)r^4\bigg] \,,
    \end{split}
    \end{equation}
    and reduces to 
    \eqref{Qdegen} when one takes into account  one of the degeneracy conditions, namely  $3B_3+2B_2 = 0$.}
    \begin{equation}
    \label{Qdegen}
        Q = \sqrt{\frac{f}{h}}\left[(F_2+2XA_1)r^4+2f\psi'XB_2r^3-\frac{1}{2}f\psi'X'(F_{3X}+4XB_6-2B_2)r^4\right] \,,
    \end{equation}
    where a subscript $X$ denotes a partial derivative with respect to $X$ and a prime a derivative with respect to $r$. Note that only two quadratic functions, $F_2$ and $A_1$, and only three cubic functions, $F_3$, $B_2$ and $B_6$ enter in the above expression, which is also the case for the equations governing axial perturbations in DHOST theories \cite{Langlois:2021aji,Langlois:2022ulw}.

    The simplicity of the equation for $\omega$,
    \begin{eqnarray}
\label{simpleequationomega}
        \frac{\mathrm{d}}{\mathrm{d}r}\bigg[Q(r)\omega'(r)\bigg] \; = \; 0 \, ,
    \end{eqnarray}
can be understood in light of some symmetry arguments, as will be explained in the next section (\ref{shiftsymmetryargument}) where we will also provide details to compute $Q(r)$. 
    The integration of \eqref{eom} leads to the expression
    \begin{equation}
    \label{omegaformula}
        \omega(r)= k \int \frac{\mathrm{d}r}{Q(r)} \, ,
    \end{equation}
    where $k$ is an integration constant, 
    proportional 
    to the angular momentum $J$.

In Horndeski theories \cite{Horndeski:1974wa}, where 
$A_1= - F_{2X}$, $2B_2 = F_{3X}$ and $B_6=0$, the expression of $Q$ reduces to
    \begin{equation}
    \label{Qhorn}
        Q = \sqrt{\frac{\f}{\h}}\left[(F_2-2XF_{2X})r^4 + \f\psi'XF_{3X}r^3\right] \, .
    \end{equation}
In General Relativity, where the only non-vanishing DHOST function is $F_2=1$, the expression of $Q$ simplifies drastically and reduces to $Q=r^4$. Therefore, we recover  \eqref{omegakerr} with $k=-6J$. 

    Let us remark that slowly rotating black holes are closely related to the study of linear perturbations about static and spherically symmetric solutions, as the Hartle-Thorne metric \eqref{hartlethornemetric} can be seen as an odd dipolar perturbation to the static black hole. Such perturbations have been studied in \cite{Kobayashi:2012kh,Takahashi:2019oxz,Cisterna:2015uya} for quadratic DHOST theories and their results are fully consistent with what we have obtained here. 

\subsection{Integrability of the equation from a conservation law}
\label{shiftsymmetryargument}
    
    \hspace{\parindent} Let us now try to understand why equation \eqref{simpleequationomega} is so simple. 
For that purpose, we first notice that under a change of the angular coordinates $\varphi \rightarrow \tilde{\varphi} = {\varphi} + \omega_0t$ where $\omega_0$ is a constant, the Hartle-Thorne metric \eqref{hartlethornemetric} becomes,
    \begin{equation}
    \begin{split}
        \mathrm{d}s^2 
        =& -\Big[ \h(r)-r^2 (\omega_0^2+2\omega_0\omega(r)) \sin^2(\theta)\Big]\mathrm{d}t^2+\frac{\mathrm{d}r^2}{\f(r)}\\
        &+r^2 \left[ \mathrm{d}\theta^2+\sin^2(\theta)\, \mathrm{d}\tilde{\varphi}^2 -2(\omega(r)+\omega_0)\sin^2(\theta) \, \mathrm{d}t\mathrm{d}\tilde{\varphi} \right] \, .
    \end{split}
    \end{equation}
    At first order in $\omega(r)$ and $\omega_0$, the line element above reduces to the original Hartle-Thorne line element where the function $\omega$ has been shifted by a constant, i.e. $\omega(r)\rightarrow\omega(r)+\omega_0$. As a consequence, from the invariance under diffeomorphisms, this implies that if $\omega(r)$ is a solution of the linearized equations, then $\omega(r)+\omega_0$ is also a solution, because they are 
    related to the same metric written in two different coordinate systems (at linear order in $\omega$). This corresponds to a continuous symmetry in the space of solutions which can be exploited in order to obtain a ``conserved'' (i.e. a constant) quantity  in the theory, which could be useful to integrate some of the equations. 
    
    However, at this stage, this symmetry is only true at linear order. Interestingly,  we can make it exact  by changing the initial Hartle-Thorne ansatz for the metric. Indeed, let us consider instead the following line element
    \begin{equation}
    \begin{split}
    \label{ansatzquad}
        \mathrm{d}s^2=&-\h(r)\mathrm{d}t^2+\frac{\mathrm{d}r^2}{\f(r)}+r^2 \left[\mathrm{d}\theta^2+\sin^2(\theta)(\mathrm{d}\varphi-\omega(r)\mathrm{d}t)^2 \right]\\
        =&-\left(\h(r)-r^2\sin^2(\theta)\omega(r)^2\right)\mathrm{d}t^2+\frac{\mathrm{d}r^2}{\f(r)}+r^2 \left[\mathrm{d}\theta^2+\sin^2(\theta)\mathrm{d}\varphi^2-2\omega(r)\sin^2(\theta)\mathrm{d}t\mathrm{d}\varphi \right]
    \end{split}
    \end{equation}
    which differs from \eqref{hartlethornemetric} by terms\footnote{It must be stressed that the quadratic term here does not have any physical meaning. 
    It has been added only to make the symmetry mentioned above exact.
    If one expands a rotating black hole metric up to quadratic terms in the angular momentum, one does not get  generally  a metric of this form.  However, since we are  only taking into account  linear terms in $\omega$ at the end,  the quadratic term  does not affect our physical results.} quadratic in $\omega$. Therefore, the equations of motion are unchanged at the linear level in $\omega$, which means that the two  ans\"atze lead to the same slowly rotating solutions.
    Furthermore, now, the change of the angular coordinate $\varphi\rightarrow \tilde\varphi=\varphi + \omega_0t$ is exactly equivalent to the shift $\omega(r)\rightarrow\omega(r)+\omega_0$. 
    As we now show,
    the conservation equation \eqref{simpleequationomega} is directly linked to this ``shift-symmetry'' (not to be confused with the shift-symmetry of the scalar field in the DHOST action).

    First of all, substituting the ansatz  \eqref{ansatzquad} into the action and computing the equation of motion for $\omega$, we get
    \begin{equation}
    \label{dSdweom}
        \frac{\delta S}{\delta\omega} = \frac{\delta S}{\delta g^{\mu\nu}}\frac{\delta g^{\mu\nu}}{\delta \omega} = -\frac{2\sqrt{-g}}{h}(\mathcal{E}_{t\varphi}+\omega\mathcal{E}_{\varphi\varphi}) \, ,
    \end{equation}
    where we have used the explicit expression of the inverse metric.
    We recognize here the combination \eqref{eom} 
    which leads directly to an integral expression for $\omega$.

    Then an immediate consequence of the symmetry mentioned above is that the Lagrangian density $\cal L$  does not depend directly on $\omega$ but on its first and second derivatives. Hence, the equation for $\omega$ is necessarily proportional to the total derivative of a quantity $\mathcal{J}$,  
    \begin{eqnarray}
    \label{conservationequation}
       \frac{\delta S}{\delta\omega} = -\frac{\mathrm{d} \mathcal{J}}{\mathrm{d}r} \, , \qquad \mathcal{J}= \frac{\partial \cal L}{\partial \omega'} - \frac{\mathrm{d}}{\mathrm{d}r} \left( \frac{\partial \cal L}{\partial \omega''}\right) \, .
    \end{eqnarray}
    Finally, a straightforward calculation leads to
    \begin{equation}
        \mathcal{J}=\sin^3(\theta)Q(r)\omega'(r)+\mathcal{O} (\omega^3) \, ,
    \end{equation}
    where we have used the explicit expression of the determinant of the metric and $Q(r)$ stands for the expression  given in \eqref{Qdegen}. In summary, the integrability of the equation for $\omega$ is directly linked to a conservation law associated with the  symmetry identified above.

\section{Shift-symmetric theories}
\label{shiftsymmetry}
\hspace{\parindent} This section is devoted to study properties of slowly rotating solutions  in shift symmetric DHOST theories. We recall that, in these theories, the functions entering in the action \eqref{DHOSTaction} depend on the kinetic density $X$ only.
We start with a study of disformal transformations of slowly rotating metrics. Then, we explicitly compute $\omega(r)$ in the case of quadratic DHOST theories. 

\subsection{Disformal Transformations}
\label{disformal}
\hspace{\parindent} In this section, we 
%study 
consider
disformal transformations, 
\begin{equation}
    \label{disform2}
     g_{\mu\nu} \; \longrightarrow \;   \tilde{g}_{\mu\nu} = C(\phi,X)g_{\mu\nu}+D(\phi,X)\,\partial_\mu\phi\, \partial_\nu\phi \, ,
    \end{equation}
where $C$ and $D$ are in general functions of $\phi$ and $X$, constrained by  the  requirement  that the disformal transformation is invertible and preserves the signature of the metric, which implies that $C>0$ and $C-2X D >0$.
For shift-symmetric theories, the functions $C$ and $D$ are 
supposed to depend on $X$ only.

Disformal transformations induce transformations in the space of DHOST theories as follows:
\begin{equation}
    \tilde{S}[\tilde g_{\mu\nu},\phi ] = S[g_{\mu\nu}, \phi] \, ,
\end{equation}
which means that, given a theory $\tilde S$ for the metric $\tilde{g}$, one defines a new theory $S$ for the metric $g$. When the disformal transformation is invertible, the two theories are classically equivalent in vacuum only while they differ when one takes into account (generic) coupling to matter. 
In particular, their causal structure is different since light geodesics in one metric do not coincide with the light geodesics in the other metric, unless the transformation is purely conformal, i.e. $D=0$. 

Let us now examine what the approximate slow-rotating metric becomes in a disformal transformation.  We start with a DHOST action $S$ and we consider a slowly rotating metric characterised by the functions $\h(r)$, $\f(r)$ and $\omega(r)$. We have shown above that $\omega$ satisfies the simple differential equation,
\begin{equation}
\label{eq_omega}
    \frac{\mathrm{d} \omega}{\mathrm{d}r} \; = \; \frac{k}{Q(r)} \, ,
\end{equation}
where $k$ is an integration constant and $Q(r)$ has been introduced in \eqref{Qdegen}. 
Performing a disformal transformation of the metric \eqref{hartlethornemetric},
we find that the new metric $\tilde{g}_{\mu\nu}$ can be written (at linear order in $\omega$) in the   form \eqref{hartlethornemetric}
\begin{equation}
    \mathrm{d}\tilde{s}^2 = - \tilde{\h} \, \mathrm{d}\tilde{t}^2 + \frac{\mathrm{d} \tilde{r}^2}{\tilde{\f}} + \tilde{r}^2 \, \mathrm{d}\Omega^2 - 2 \tilde r^2 \tilde{\omega}\sin^2 \theta \, \mathrm{d}\tilde{t} \, \mathrm{d} \tilde{\varphi} \, + {\cal O}(\omega^2) \, ,
\end{equation}
when using  new coordinates $(\tilde{t},\tilde{r},\tilde{\varphi})$  defined by 
\begin{eqnarray}
\label{tilde_coordinates}
     \mathrm{d}\tilde{t} = \mathrm{d}t - \frac{qD(r)\psi'(r)}{C(r) h(r)-D(r)q^2}\mathrm{d}r \, , \quad 
        \mathrm{d}\tilde{\varphi} = \mathrm{d}\varphi - \frac{q D(r)\psi'(r)\omega(r)}{C(r) h(r)-D(r)q^2}\mathrm{d}r \, , \quad \tilde{r}=R(r)\equiv r\sqrt{C(r)}  \,.
\end{eqnarray}
The new metric components are related to the old ones  via the expressions
\begin{eqnarray}
\label{tilde_functions}
    \tilde \h(\tilde r) = C \h - q^2 D \, , \qquad
    \frac{1}{\tilde \f(\tilde r)} = \frac{C}{R'^2}\left(\frac{1}{\f} + \frac{D \h \psi'^2}{C\h-q^2 D}\right) \, , \qquad \tilde\omega(\tilde r) = \omega \, ,
\end{eqnarray}
where the functions of the r.h.s. of the two identities are evaluated on $r$, viewed as a function of $\tilde{r}$.
Let us make two comments: we have used the notation $C(r)$ for $C(X(r))$ and $D(r)$ for $D(X(r))$; as $C$ is supposed to be positive, the coordinate $\tilde{r}$ is well-defined. We can formally integrate the two differential equations in \eqref{tilde_coordinates} so that
\begin{eqnarray}
    \tilde{t} = t + \mu(r) \, , \qquad \tilde{\varphi} = \varphi + \nu(r) \, ,
\end{eqnarray}
where $\mu$ and $\nu$ are given in terms of integrals.
As a result, $\tilde{\omega}$ satisfies an equation similar to \eqref{eq_omega},
\begin{equation}
\label{eq_omegatilde}
    \frac{\mathrm{d} \tilde{\omega}}{\mathrm{d}\tilde r} \; = \; \frac{{k}}{\tilde Q(r)} \, ,
\end{equation}
where  $\tilde{Q}$ is the same function as in \eqref{Qdegen} but expressed in terms of the new functions $\tilde{F}_i$, $\tilde{A}_i$ and $\tilde{B}_i$ characterising the action $\tilde{S}$.
Of course, the two equations \eqref{eq_omega} and \eqref{eq_omegatilde} must be equivalent, which leads to the consistency relation
\begin{equation}
\label{conditiondis}
    R' \frac{Q}{\tilde{Q}} = 1 \,.
\end{equation}
This relation can be verified explicitly by using the expressions of $\tilde{F}_i$, $\tilde{A}_i$ and $\tilde{B}_i$ given in terms of $F_i, A_i$ and $B_i$, as we show in Appendix \ref{disformalappendix}.

In summary, 
given a DHOST action $S$ that admits a slowly rotating solution associated with the function $\f$, $\h$ and $\omega$, then the disformally related action $\tilde S$ admits also a slowly rotating solution associated to $\tilde \f$, $\tilde \h$ and $\tilde \omega$ related to the original functions by \eqref{tilde_functions}. In particular, the frame dragging functions are related by
\begin{equation}
\label{omega_disformal_relation}
    \omega(r)=\tilde{\omega}\left( R(r)\right) \, , \quad \text{with} \quad R(r) = r\sqrt{C(r)} \, .
\end{equation}
Expressing $\tilde{\omega}$ in terms of $\omega$ can be done only if the coordinate transformation $\tilde r= \sqrt{C(r)}r$ is invertible, or equivalently $\tilde r$ is a monotonous function of $r$. 
    
\subsection{Quadratic theories}
\label{quadratictheories}

  \hspace{\parindent} In this subsection, we focus on quadratic shift-symmetric DHOST theories \eqref{DHOSTaction}, which can be related to quadratic Horndeski theories via disformal transformations.

  For quadratic 
  Horndeski theories, the expression of $Q$ \eqref{Qhorn} reduces to the simple form
    \begin{equation}
    \label{Q_quad_Horn}
        Q = \sqrt{\frac{\f}{\h}}(F_2-2XF_{2X})r^4 \, .
    \end{equation}
    This expression can be further simplified by taking into account 
    the equations of motion, as we now show. Due to the shift-symmetry of the theory, the equation of motion for the scalar field is a conservation equation, which means that it can be written as the divergence of a Noether current $J^\mu$\cite{Babichev:2013cya},
    \begin{equation}
        \mathcal{E}_\phi = \frac{1}{\sqrt{-g}}\frac{\delta S}{\delta\phi} = 0 \; \Longleftrightarrow \; \nabla_\mu J^\mu = 0\, .
    \end{equation}
    Furthermore, it was shown in \cite{Babichev:2015rva} that $\mathcal{E}_t^r=qJ^r$, which necessarily implies $J^r =0$ when $q\neq0$. In the case where $q=0$, an argument using the physical necessity to keep $J_\mu J^\mu$ finite on the horizon also gives $J^r=0$ (see \cite{Hui:2012qt} for instance). 
    
    Finally, the  combination 
    \begin{equation}
        \mathcal{E}_{rr}+\frac{1}{\f\h}\mathcal{E}_{tt}-\frac{1}{\f}\left(\psi'+\frac{q^2}{\f\h\psi'}\right)J^r=-\frac{2}{r}\sqrt{\frac{\h}{\f}}~\frac{\mathrm{d}}{\mathrm{d}r}\left(\sqrt{\frac{\f}{\h}}(F_2-2XF_{2X})\right)=0 \, ,
    \end{equation}
    shows that the coefficient in front of $r^4$ in the expression \eqref{Q_quad_Horn} of $Q$ is constant. 
    As a consequence, $Q$ is proportional to $r^4$, as in General Relativity, and 
    the function $\omega(r)$ is  identical to the one of Kerr \eqref{omegakerr} for all quadratic shift-symmetric Horndeski theories (even when $\f(r)$ and $\h(r)$ are different from the Schwarzschild metric coefficients). Hence, 
    we have
    \begin{equation}
    \label{omega_Horn}
        \omega(r) = \frac{2J}{r^3} \ .
    \end{equation}
    \par This result can be 
    extrapolated
    to other quadratic shift-symmetric DHOST theories by resorting to disformal transformations.
    Indeed, given a  slowly rotating metric $\tilde{g}_{\mu\nu}$ defined in  a Horndeski theory  $\tilde{S}$, the slowly rotating metric 
   $g_{\mu\nu}$ in the DHOST theory $S$ verifies
   \begin{eqnarray}
       \omega(r) = k \int \frac{\sqrt{\h/\f}} {(F_2 + 2X A_1)r^4} \mathrm{d}r \; = \;\frac{2J}{ r^3}   C(r)^{-3/2} \, ,
   \end{eqnarray}
   according to \eqref{omegaformula} and \eqref{Qdegen} for the first equality, and to \eqref{omega_disformal_relation} and \eqref{omega_Horn} for the second equality.
Notice that all the functions entering into the integral are viewed as functions of $r$ and the integration constant has been fixed to zero.

    In particular, the expression for $\omega$
    is the same as in General Relativity for shift-symmetric quadratic beyond Horndeski theories \cite{Gleyzes:2014dya} which are related to Horndeski theories by a pure disformal transformation (i.e. with $C=1$).  This is worth noting because many static solutions were found in this framework, see for instance \cite{Rinaldi:2012vy,Babichev:2013cya,Babichev:2017guv,Bergliaffa:2021diw,Bakopoulos:2023fmv,Charmousis:2025xug}.
    
    \par Otherwise, the function $\omega$ is different from the Kerr expression. Nonetheless, if we want the disformal transformation to preserve the asymptotic flatness of the metric, we must impose that $C = 1 + o(1)$ at infinity (when $r \rightarrow \infty$). In this case, the asymptotic behaviour of $\omega(r)$ remains the same as in Kerr.

\subsection{Circular time-like and light-like geodesics}
\label{primaryhair}

\hspace{\parindent} As an application of the general formalism developed above, we consider the exact static black holes obtained recently \cite{Bakopoulos:2023fmv}, in the context of   specific  shift-symmetric quadratic DHOST theories defined by the non-trivial functions
\begin{equation}
\label{theory_primary_hair}
    F_0= -\frac{2\alpha}{\lambda^2}\,X^p, \quad F_2=1-2X A_1, \quad A_1=-A_2 = \frac{\alpha}{2}\,X^{p-1}\,, \quad 
    A_3=-A_4=\frac{\alpha}{2}(2p-1)X^{p-2}\,, %\quad A_5=0 \,,
\end{equation}
and $A_5=0$,
where $\alpha$ is a coupling constant and $p$  some constant, conveniently chosen to be an integer or half-integer. Notice that these theories belong to the quadratic Beyond Horndeski theories.

The black holes obtained in these theories  are interesting as they are the first solutions in DHOST theories characterized by a primary hair, i.e. a scalar charge that can be fixed  independently of the mass of the black hole. This scalar parameter $q$ is associated with the linear  time dependence of the scalar field, namely
\begin{equation}
\phi(r,t)= qt+ \psi(r)\,,
\end{equation}
where the radial dependent part $\psi(r)$ satisfies 
\begin{equation}
\psi^{\prime}(r)^2=\frac{q^2}{\cA(r)^2}\left[1-\frac{\cA(r)}{1+(r / \lambda)^2}\right] \,, \label{scalar}
\end{equation}
so that  the scalar kinetic term is given by\footnote{The solution is asymptotically characterized by a null scalar gradient with $X\to 0$ in a Minkowski space-time.  Hence, perturbations of this solution may be subtle with possible strong-coupling issues for scalar fluctuations. This point deserves further  investigation that we hope to pursue in the future. We thank the referee for observation.}
\begin{equation}
X=\frac{q^2 /2}{1+(r / \lambda)^2}\,,
\end{equation}
which is regular and decays when going away from the black hole.

The explicit black hole metric depends on the value of $p$ that defines the DHOST theory \eqref{theory_primary_hair}. For $p=1/2$, the metric is simply Schwarzschild, thus defining a stealth solution with a non trivial scalar profile. An interesting value is $p=2$, for which the metric is defined by 
\begin{equation}
\label{A_n2}
\cA(r)=\cB(r)=1-\frac{2 M}{r}+\xi_2\left(\frac{\pi / 2-\arctan (r/\lambda)}{r/\lambda }+\frac{1}{1+(r/\lambda)^2}\right) \,, \qquad \xi_2\equiv \frac34 \alpha q^4\,.
\end{equation}
From our discussion in the previous subsection,  we know that slow-rotating  version of these solutions, i.e. of the form \eqref{hartlethornemetric}, are such that $\omega(r)$ is  the same as in General Relativity.

It is then instructive to study time-like and light-like geodesics in the equatorial plane $\theta=\pi/2$. As a consequence of the time and angular invariance of the metric, one can introduce as usual  the conserved energy and angular momentum
\begin{equation}
E=\cA \dot t +\omega r^2 \dot\varphi\,, \qquad L=r^2 \dot\varphi-\omega r^2 \dot t\,,
\end{equation}
where a dot denotes a derivative with respect to the geodesic parameter (the proper time $\tau$ for a time-like geodesic and an affine parameter $\lambda$ for a light-like geodesic). Inverting the above system to obtain  $\dot t$ and $\dot\varphi$ in terms of $E$ and $L$, and substituting in the expression $g_{\mu\nu}\dot x^\mu \dot x^\nu=-\kappa$ (with $\kappa=1$ for time-like geodesics and $\kappa=0$ for light-like ones), one gets, at first order in $\omega$, the radial equation
\begin{equation}
\frac12 \dot r^2+V_{\rm eff}(r)=0\,,
\end{equation}
with the effective potential
\begin{equation}
V_{\rm eff}(r)=\frac{L^2 \cB}{2r^2}-\frac{E^2\cB}{2\cA}+\frac{EL\cB}{\cA}\omega+\frac{\kappa}{2}\cB\,.
\end{equation}
For massive particles,  the innermost stable circular orbit is defined by the three conditions
\begin{equation}
{\rm ISCO:}\qquad V_{\rm eff}(r)=0\,,\qquad V'_{\rm eff}(r)=0\,,\qquad V''_{\rm eff}(r)=0\qquad ({\rm with}\  \kappa=1)\,.
\end{equation}
The first two conditions define a circular orbit whereas the last one characterizes the onset of instability, i.e. the transition from stable circular orbits with $V''_{\rm eff}(r)>0$ to unstable ones with $V''_{\rm eff}(r)<0$.

\begin{figure}[h]
\begin{center}
\includegraphics[width=10cm]{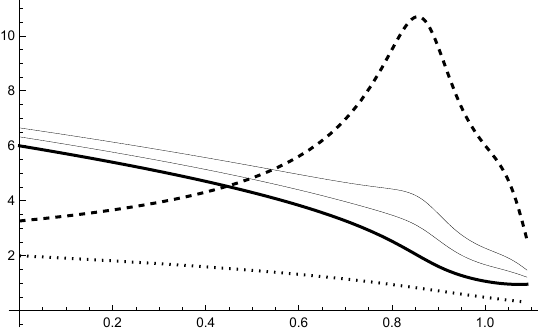}
\end{center}
\caption{Plot of the non-rotating ISCO (thick continuous line) as a function of the parameter $\xi_2$  compared with the BH horizon (dotted line). Both radii are expressed in units of the BH mass $M$. Above the non-rotating ISCO, two thin lines indicate the modified ISCO for $a=0.1$ and $a=0.2$ (in the linear approximation) for retrograde orbits. More generally, the derivative $\partial r^-_{\rm ISCO}/\partial a$ is plotted with a dashed line. For all plots, we take $\lambda=M$.}
\label{fig1}
\end{figure}

As a first step, one can determine the ISCO as a function of $\xi_2$, for the non-rotating metric (i.e. J=0) \eqref{A_n2}. The result is plotted on Fig.~\ref{fig1} , together with the position of the BH horizon, which also depends on the parameter $\xi_2$ (for the choice of parameter $\lambda=M$). We then compute the ISCO when a slow rotation, parametrized by $a=J/M$, is introduced.
In the limit $\xi_2 \rightarrow 0$, one recovers the well-know results for Schwarzschild $r_{\rm ISCO}=6M$ and for Kerr in the slow rotation limit:
\begin{equation}
    \frac{\partial r^-_{\rm ISCO}}{\partial a}(\xi_2=0)=4\sqrt{\frac{2}{3}}\,,
\end{equation}
{where $r^-_{\rm ISCO}$ denotes here the ISCO for retrograde orbits, i.e. orbiting the BH in the sense opposite to the BH rotation. The analogous expression for prograde orbits has the opposite sign. In other words, the ISCO radius increases with the BH rotation for retrograde orbits but decreases for prograde orbits.
One can notice that the position of the ISCO is most sensitive to the rotation when $\xi_2\simeq 0.9$, for the particular choice $\lambda=M$, as illustrated on Fig.~\ref{fig1}.
}

\begin{figure}[ht]
\begin{center}
\includegraphics[width=10cm]{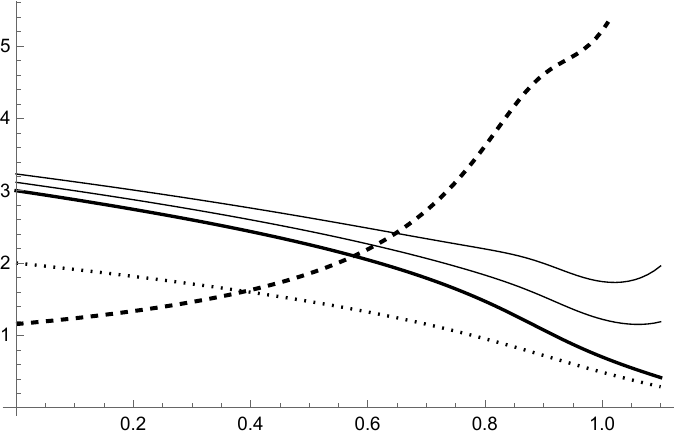}
\end{center}
\caption{Plot of the 
light circular orbit radius $r_{\rm LCO}$
 in the non-rotating solution (thick continuous line), as a function of the parameter $\xi_2$  compared with the BH horizon (dotted line). Both radii are expressed in units of the BH mass $M$. Above the non-rotating 
 case, two thin lines indicate the modified %photon ring 
 light radius
 for $a=0.1$ and $a=0.2$ (in the linear approximation) for retrograde orbits. More generally, the derivative $\partial r^-_{\rm LCO}/\partial a$ is plotted with a dashed line. For all plots, we take $\lambda=M$.}
\label{fig2}
\end{figure}

Let us now turn to massless particles and their circular geodesics in the equatorial plane (which are unstable). 
The radial position of these  orbits is obtained by solving 
\begin{equation}
V_{\rm eff}(r)=0\,,\qquad V'_{\rm eff}(r)=0\,,\qquad  ({\rm with}\  \kappa=0)\,.
\end{equation}
The analysis of the position of the circular light orbits is very similar to that of the ISCO and we have plotted the results on Fig.~\ref{fig2}. As expected, we see that the radius of the photon orbits varies with the deformation parameter $\xi_2$ and with the rotation (as in GR). We also illustrate the variation of the retrograde orbits $r_{\rm LCO}^-$ which increases with the black hole angular momentum. 

\section{On the angular dependence of the frame dragging function}
\label{angulardep}
\hspace{\parindent} 
So far, we have made the assumption that the frame dragging function $\omega$ depends only on the radial coordinate $r$. In the context of General Relativity, such an assumption is well-justified because of the no-hair theorem \cite{Carter:1971zc} which ensures that the slowly rotating solution  must correspond to the small angular momentum limit of the Kerr solution. Thus, there is no reason to expect $\omega$ to depend on the angular coordinate $\theta$. 

This argument is however no longer valid in modified theories of gravity, which admit hairy black hole solutions. Therefore, we study in this section the possibility that $\omega$ depends on $\theta$. We start by computing the equation of motion for $\omega$. Then, we revisit the case of General Relativity and recall that regularity arguments prevent $\omega$ to depend on $\theta$. Finally, we adapt the analysis to the specific case of  DHOST theories and argue that $\omega$ should not depend on $\theta$.

\subsection{Equation of motion}
    \hspace{\parindent} 
    Assuming a $\theta$-dependence for $\omega$  leads to a  partial differential equation, instead of the ordinary differential equation  \eqref{eom}. 
    In order to compute this  equation, we proceed as in section \ref{shiftsymmetryargument} since 
    the shift-symmetry argument remains valid 
    and finally obtain (see Appendix \ref{wrthappendix} for more details)
    \begin{equation}
    \label{eqwrth}
        \frac{\partial}{\partial r}\left[Q(r)\sin^3(\theta)\frac{\partial\omega}{\partial r}\right]+\frac{\partial}{\partial\theta}\left[\Qa(r)\sin^3(\theta)\frac{\partial\omega}{\partial\theta}\right]=0 \, ,
    \end{equation}
    where $Q$ is the same function as in \eqref{Qdegen} while $\Qa$ is defined by
    \begin{equation}
    \begin{split}
    \label{Qtheta}
        \Qa(r)=\frac{r^2}{\sqrt{\h\f}}\bigg[&\left(F_2+\frac{q^2}{\h}A_1\right)+
        %f\psi'\frac{q^2}{h}(3B_3+2B_2)r+
        \frac{1}{2}\f\psi'X'\left(F_{3X}-2\frac{q^2}{\h}B_6\right)-\frac{q^2}{\h}B_2\frac{(\h X)'}{\h\psi'}\bigg] \, ,
    \end{split}
    \end{equation}
    for DHOST theories\footnote{ Similarly to $Q$, the expression \eqref{Qtheta} for $\Qa$ is obtained from
    \begin{equation*}
          \Qa(r)=
          \frac{1}{\sqrt{\h\f}}\bigg[\left(F_2+\frac{q^2}{\h}A_1\right)r^2+
        \f\psi'\frac{q^2}{\h}(3B_3+2B_2)r+
        \frac{1}{2}\f\psi'X'\left(F_{3X}-2\frac{q^2}{\h}B_6\right)r^2-\frac{q^2}{\h}B_2\frac{(\h X)'}{\h\psi'}r^2\bigg] \,,
    \end{equation*}
    after imposing the degeneracy condition $3B_3+2B_2=0$.}.
    
    To solve \eqref{eqwrth}, 
    we proceed as in \cite{Hartle:1967he} and decompose $\omega(r,\theta)$ on a basis of Legendre polynomials\footnote{The coefficient of $\mathrm{d}t\mathrm{d}\varphi$ in the line element, which involves $\omega$, can be seen as an axial (or odd-parity) perturbation of the static and spherically symmetric background. Hence, it can be decomposed into the following basis of vectors on the 2-sphere,
$$        V_{(l,m)}^a = %\epsilon^{ab}\partial_bY_l^m =
        \left(\frac{1}{\sin(\theta)}\frac{\partial Y_l^m}{\partial\varphi};-\sin(\theta)\frac{\partial Y_l^m}{\partial\theta}\right) \, .
    $$
    Since we are considering an axisymmetric metric (whose coefficients do not depend on $\varphi$), only the modes $m=0$ enter in the decomposition, hence  
    \begin{equation*}
        \omega(r,\theta)\sin^2(\theta)=\sum_{l=1}^{+\infty}\omega_l(r)V_{(l,0)}^\varphi(\theta) = -\sum_{l=1}^{+\infty}\omega_l(r)\sin(\theta)\frac{\partial P_l}{\partial\theta} \; .
    \end{equation*}},
    \begin{equation}
    \label{decomposition}
        \omega(r,\theta)\sin^2(\theta)=
        -\sum_{l=1}^{+\infty}\omega_l(r)\sin(\theta)\frac{\partial P_l}{\partial\theta} \,,
    \end{equation}
        where $P_l$ is the Legendre polynomial of order $l$ evaluated at $\cos(\theta)$. Substituting \eqref{decomposition} into \eqref{eqwrth} leads to the following ordinary differential equation for each mode $\omega_l$:
    \begin{equation}
    \label{eqdec}
        \frac{\mathrm{d}}{\mathrm{d}r}\left(Q\,\frac{\mathrm{d}\omega_l}{\mathrm{d}r}\right)+\big(2-l(l+1)\big)\Qa\, \omega_l = 0 \, .
    \end{equation}
    For $l=1$, the second term in \eqref{eqdec} vanishes and we recover the solution  \eqref{omegaformula}  discussed in the previous sections. For $l\geq2$, 
    the analysis depends on the form of the functions $Q$ and $\Qa$. In the next three subsections, we discuss in turn three specific cases: General Relativity, quadratic DHOST theories of the form  \eqref{theory_primary_hair} and 4D-Einstein-Gauss-Bonnet (which is another particular case of DHOST theories).

\subsection{The case of General Relativity revisited}
    \hspace{\parindent} 
As we mentioned above, the no-hair theorem \cite{Carter:1971zc} in General Relativity ensures that $\omega$ does not depend on $\theta$. 
     Interestingly, one can recover this result from the differential equation \eqref{eqdec}, as discussed in \cite{Pani:2009wy}. 

\medskip

    In General Relativity, the functions $Q$ and $\Qa$ in \eqref{eqdec} reduce to
    \begin{eqnarray}
        Q(r)=r^4 \, , \qquad \Qa=\frac{r^2}{f(r)} \, , \qquad f(r)= 1-\frac{2M}{r} \, .
    \end{eqnarray}
    Expressing, for simplicity, the radial coordinate $r$ in units of the Schwarzschild radius, which is equivalent to taking $2M=1$, equation \eqref{eqdec} reduces to
    \begin{equation}
    \label{eqgr}
        \left(1-\frac{1}{r}\right)\frac{\mathrm{d}}{\mathrm{d}r}\left(r^4 \omega_l'\right)+\big(2-l(l+1)\big)r^2\omega_l=0 \, .
    \end{equation}
    For $l=1$, we obtain the solution corresponding to the Kerr metric \eqref{omegakerr} as expected. For $l \geq 2$, the solution is  more involved but can be written in terms of special functions \cite{Pani:2009wy}. 
    
    For the present discussion,  we simply need   to study the asymptotic behaviour, at spatial infinity and near the horizon, of $\omega_l(r)$ or, equivalently, of the function $y_l(r)=r^2\omega_l(r)$, which satisfies the simpler differential equation
    \begin{equation}
    \label{eqgry}
        r(r-1) \, y_l''+\left(\frac{2}{r}-l(l+1)\right)\, y_l \, = \, 0 \, .
    \end{equation}
    The general solution of \eqref{eqgry} can be written as
    \begin{eqnarray}
    \label{gensol}
        y_l(r) \; = \; c_{(1)} \, y_l^{(1)}(r) + c_{(2)} \, y_l^{(2)}(r)  \, , 
    \end{eqnarray}
    where $(y_l^{(1)},\, y_l^{(2)})$ is some basis of solutions while $c_{(1)}$ and $c_{(2)}$ are constants (to be fixed with boundary conditions). 
For $l=2$,  the general solution is given explicitly by
    \begin{equation}
        y_2(r)=c_{(1)} \, r^2(r-1)+ 5 c_{(2)} \, \left(4r^2-2r-\frac{2}{3}-\frac{1}{3r}+4r^2(r-1)\ln\left(\frac{r-1}{r}\right)\right).
    \end{equation}
    
    At spatial infinity, the leading order of the two independent solutions $y_l^{(1)}$ and $ y_l^{(2)}$ can be deduced from  the equation
    \begin{equation}
    \label{asymptoticeqGR}
        r^2 \, y_l''-l(l+1)\, y_l \, {=} \, 0 \, ,
    \end{equation}
     where the coefficients in  \eqref{eqgry}  have  been replaced by their leading order expression at large $r$ (see e.g. chapter 4.2 of \cite{Teschl2012} for details). 
    Since the general solution of \eqref{asymptoticeqGR} is simply given by a linear combination of $r^{l+1}$ and $r^{-l}$, we can choose a basis $(y_l^{(1)},\, y_l^{(2)})$ such that
    \begin{equation}
    \label{asymptoticsolGR}
        y_l^{(1)}(r) = r^{l+1} (1+o(1)) \, , \qquad
        y_l^{(2)}(r) = r^{-l} (1+o(1)) \, .
    \end{equation}
    As the solution should not diverge at spatial infinity in  order to recover an asymptotically Minkowski metric,  we have to fix $c_{(1)}=0$ for every $l$ in \eqref{gensol}. 
    
    Let us now turn to the asymptotic behaviour at the horizon.  In the case $l=2$, one sees that the solution $y_l^{(2)}$ is pathological at the horizon ($r \rightarrow 1$)
    because of the logarithm term: the solution is
    continuous but not differentiable. As a consequence, $c_{(2)}$ must also vanish if one imposes that the metric components is both continuous and differentiable. 
    The same conclusion applies for $l>2$, due to the presence of a similar logarithmic term leading to a pathological behaviour  at the horizon (see \cite{Pani:2009wy} for instance).  Hence, all modes $l\geq2$ must vanish and $\omega$ cannot depend on $\theta$, as expected from the no-hair theorem. 
    
\subsection{Generalization to a class of quadratic DHOST theories}
    \hspace{\parindent} We now extend the analysis of the previous subsection to  the family of quadratic  shift-symmetric DHOST theories considered in section \ref{primaryhair}, which admit interesting hairy black hole solutions \cite{Bakopoulos:2023fmv} where the scalar field is of the form $\phi=qt+\psi(r)$ while the metric is static and homogeneous, i.e. $f(r)=h(r)$. 
    
    We can easily compute the two functions entering the equation for $\omega$ and we find $Q(r)=r^4$, as in General Relativity, while  $\Qa$ is given by
    \begin{equation}
        \Qa=\frac{r^2}{f}\left(F_2+\frac{q^2}{f}A_1\right) 
        %= \frac{r^2}{f}\left(F_2+2XA_1+f\psi'^2A_1\right)
        =\frac{r^2}{f}\left(1+f\psi'^2A_1\right) \, .
    \end{equation}
    The equation for the  function  $y_l(r)=r^2\omega_l(r)$ now reads
    \begin{equation}
    \label{eqy1}
     \alpha_2(r) \, y_l''+\alpha_0(r) \, y_l=0 \,, \qquad
        \alpha_2=r^2f \, , \quad \alpha_0=\big(2-l(l+1)\big)\left(1+f\psi'^2A_1\right)-2f \, \, .
    \end{equation}
where the coefficients behave, at spatial infinity, as
\begin{eqnarray}
    \alpha_2(r) = r^2(1+o(1)) \, , \qquad\alpha_0(r) = -l(l+1) + o(1) \, ,
\end{eqnarray}
assuming\footnote{The case $p=1$ corresponds to a non-asymptotically flat space-time.} $p >1$.
At spatial infinity, we thus recover the same asymptotic behaviour as in General Relativity, which implies that
one of the two branches of solutions has to be eliminated, say $c_{(1)}=0$. 

At the horizon, by contrast, the behaviour of the solution  differs from that in General Relativity.
Indeed, 
near the horizon radius $r_H$ (defined by $f(r_H)=0$), the two functions $\alpha_0$ and $\alpha_2$ behave as 
\begin{eqnarray}
    \alpha_2= \alpha_{2H} \varepsilon (1+ o(1))\, , \qquad
\alpha_0 = \frac{\alpha_{0H}}{\varepsilon}(1+ o(1)) \, , 
\end{eqnarray}
where $\varepsilon = r-r_H$, $X_H=X(r_H)$ and
\begin{eqnarray}
   \alpha_{2H}= r_H^2 f'(r_H) \, , \qquad
    \alpha_{0H} = \alpha q^2\frac{ X_H^{p-1}}{2f'(r_H)} (2-l(l+1)) \, .
\end{eqnarray}
Hence, one can choose a basis of solutions \eqref{gensol} of \eqref{eqy1} such that 
\begin{eqnarray}
    y_l^{(1)} = \varepsilon^{n_+} (1+o(1)) \, , \qquad
    y_l^{(2)} = \varepsilon^{n_-} (1+o(1)) \, ,
\end{eqnarray}
where the powers $n_\pm$ are given by
 \begin{equation}
        n_\pm = \frac{1}{2} \left[1\pm\left({1 - 4\frac{\alpha_{0H}}{\alpha_{2H}}
        }\right)^{1/2} \right]\, .
    \end{equation}
When $l=1$, $n_+=1$ and $n_-=0$, thus the two branches are regular (continuous and differentiable) at the horizon. Hence, as in General Relativity, the mode $\omega_1$ should be included.
For the modes $l\geq2$, we must examine the near-horizon behaviour depending on the sign of $\alpha$. 

If $\alpha>0$,  we see that $n_-<0$, giving a diverging branch at the horizon, which must be eliminated.  Provided that this branch is not the same  as the one already eliminated at spatial infinity, we conclude that $y_l=0$ for all $l\geq2$. Moreover, except for some specific values of $\alpha$, $n_+$ is not an integer and so the remaining branch at the horizon is not $\mathcal{C}^\infty$. If we demand the solution to verify this property, we can eliminate both branches at the horizon.
    
   If $\alpha < 0$,  $n_\pm$ are either real or complex. When they are real, it is easy to see that $0<n_- <n_+<1$, so both branches are continuous but not differentiable at the horizon, and must therefore be eliminated.  When $n_\pm$ are complex, the solutions are of the form,
    \begin{equation}
        y_l^{(1)}= \sqrt{\varepsilon}~e^{i\Im(n_+)\ln(\varepsilon)} (1+o(1))\, , \qquad 
        y_l^{(2)}=\sqrt{\varepsilon}~e^{-i\Im(n_+)\ln(\varepsilon)} (1+o(1)) \, ,
    \end{equation}
    where $\Im(n_\pm)$ denotes the imaginary part. 
    Both branches are continuous but not differentiable at the the horizon, so we can eliminate both again if we demand the solution to be differentiable.

 In conclusion,  $\omega$ cannot depend on the angle $\theta$.

\subsection{Four Dimensional Einstein-Gauss-Bonnet}
    \hspace{\parindent} In this section, we extend our analysis to the case of a four dimensional Einstein-Gauss-Bonnet theory\footnote{In this work, we adopt the prescription of \cite{Lu:2020iav}, in which the limiting procedure introduces an additional scalar degree of freedom. Other prescriptions exist, for instance  those that construct a consistent D→4 Einstein–Gauss–Bonnet theory (in an unique way) with only two dynamical degrees of freedom by breaking temporal diffeomorphism invariance \cite{Aoki:2020lig}. }, for which a static black hole solution is known \cite{Lu:2020iav}. A slowly rotating solution has already been found in \cite{Charmousis:2021npl}, and was studied in \cite{Gammon:2022bfu}. This theory belongs to the class of Horndeski theories, and is defined by the following non-trivial functions,
    \begin{equation}
    \label{theorygb}
    \begin{split}
        & F_0=8\alpha X^2~~,~~F_1=8\alpha X~~,~~F_2=1+4\alpha X ~~,~~A_1=-A_2=-4\alpha \, ,\\
        &F_3=-4\alpha\ln\rvert  X\lvert~~,~~6B_1=-2B_2=3B_3=\frac{4\alpha}{X} \, ,
    \end{split}
    \end{equation}
    with $\alpha$ being a coupling constant of dimension length squared. As this theory contains cubic terms in the second derivatives of the scalar field, it does not fall under the class of theories we have studied in the previous section \ref{quadratictheories}. 
    
    The static black hole solution is given by,
    \begin{equation}
        f(r)=h(r)=1+\frac{r^2}{2\alpha}\left(1-\sqrt{1+\frac{8\alpha M}{r^3}}\right) \, ,
    \end{equation}
    while  the scalar field  can be obtained from,
    \begin{equation}
        \psi'(r)=\frac{\sqrt{f(r)}-1}{r\sqrt{f(r)}} \,, 
    \end{equation}
    where, for simplicity, we restrict ourselves to the case of $q=0$. 
    
    The slowly rotating function $\omega(r)$ still satisfies the equation \eqref{eqwrth} where the two functions $Q$ and $\hat Q$ are now given by, 
    \begin{equation}
    \label{Qgb}
        Q=r^4\sqrt{1+\frac{8\alpha M}{r^3}} \, , \qquad
        \hat Q=\frac{r^2}{f}\left[1+\frac{2\alpha}{r^2}\left(f-1-rf'\right)\right] \, .
    \end{equation}
    We start by considering the $l=1$ mode, and an  immediate calculation leads to the same solution as the one obtained in \cite{Charmousis:2021npl},
    \begin{equation}
        \omega_1(r)=
        \frac{J}{2\alpha M}\left(\sqrt{1+\frac{8\alpha M}{r^3}} -1\right) \, .
    \end{equation}
    Notice that the two integration constants that show up when we integrate $\omega_1$ are chosen such that $\omega_1$ tends to the Kerr solution at spatial infinity. 
    
    In order to study the modes for $l >1$, we proceed as in the previous section. We first perform the change of variable $y_l=\sqrt{Q}~\omega_l$ so that the equation for $y_l$ does not involve a first derivative term,
    \begin{equation}
    \label{eqgb}
        y_l''+\left[\frac{Q'^2-2QQ''}{4Q^2}+\left(2-l(l+1)\right)\frac{\hat Q}{Q}\right]y_l=0 \, .
    \end{equation}
    At spatial infinity, the coefficient of $y_l$ behaves as follows,
    \begin{eqnarray}
        \frac{Q'^2-2QQ''}{4Q^2}+\left(2-l(l+1)\right)\frac{\hat Q}{Q} = -\frac{l(l+1)}{r^2}  + o\left(\frac{1}{r^2}\right) \, ,
    \end{eqnarray}
hence, the asymptotic behaviour is the same as  in General Relativity  \eqref{asymptoticeqGR} and we can eliminate the branch that diverges like like $r^{l+1}$.

    At the horizon, located at radius $r_H=M+\sqrt{M^2-\alpha}$, the function $\hat Q$ diverges (due to the presence of $f(r)$ in the denominator) while $Q$ and its derivatives are finite. Thus, the coefficient of $y_l$ behaves as follows,
    \begin{eqnarray}
        (2-l(l+1)) \frac{\alpha_H}{\varepsilon} \, , \qquad \alpha_H= \frac{r_H^4-4\alpha Mr_H}{2r_H^2(r_H-M)(r_H^2+2\alpha)} \, ,
    \end{eqnarray}
    where $\varepsilon= r-r_H$. Hence, the solution behaves as in General Relativity at the horizon with a logarithmic term $(r-r_H)\ln(r-r_H)$ appearing in one of the two branches of solutions (for $l\neq 1$), making it non-differentiable. Hence, by requiring the solution to be differentiable, we can also eliminate a branch of solution.

    We have  no  proof that the branch eliminated at infinity is not the same as the one eliminated at the horizon, in which case a regular branch of solution would remain. However, given that the theory \eqref{theorygb} reduces to GR continuously in the limit $\alpha\rightarrow0$, one can expect that the two branches eliminated at spatial infinity and at the horizon, respectively, are distinct, as in GR, at least for sufficiently small values of $\alpha$. This would confirm that $\omega$ does not depend on the angle $\theta$.

\section{Conclusion}

\hspace{\parindent}We studied slowly rotating black-hole solutions within the framework of DHOST theories. Starting from any static, spherically symmetric background of a DHOST model, we derived a general integral expression for the frame-dragging function $\omega$. The possibility to obtain such a general result relies on an interesting "hidden" symmetry of the equations of motion, which leads to their integrability.
We then showed that this expression is consistent with disformal transformations and derived how $\omega$ transforms under such transformations.
We illustrated our results with explicit examples and proved (using regularity conditions at infinity and at the horizon) that no solution for $\omega$ with angular dependence can exist, as in General Relativity. In the context of the recently discovered black holes with primary hair, we analyzed the time-like and light-like circular geodesics  and quantified how they deviate from the predictions of the Kerr solution, in the slow-rotation limit.

It would be interesting to investigate whether the method for computing slowly rotating solutions, along with the integrability property of the equations of motion, remains valid in other modified theories of gravity. We also aim to generalize our results to the case of slowly rotating stars.

\subsection*{Acknowledgements}
This work was partially supported by the French National Research Agency (ANR) via Grant No.
ANR-22-CE31-0015-01 associated with the project StronG. We thank Christos Charmousis and  Mokhtar Hassaine for insightful discussions.

\appendix
\section{Appendix: Disformal Transformations}
\label{disformalappendix}
\hspace{\parindent} In this appendix, we provide technical details on the effect of disformal transformations on slowly rotating spacetimes in DHOST theories. 

We recall that a disformal transformation of the metric is defined by,
\begin{equation}
    \label{disform2_app}
     g_{\mu\nu} \; \longrightarrow \;   \tilde{g}_{\mu\nu} = C(\phi,X)g_{\mu\nu}+D(\phi,X)\, \partial_\mu\phi\, \partial_\nu\phi \, ,
    \end{equation}
where $C$ and $D$ are functions of $\phi$ and $X$, constrained by  the  requirement  that the disformal transformation is invertible. Moreover, we assume that the conformal factor $C$ is a positive function to keep the signature of the metric unchanged. As we restrict ourselves to shift-symmetric theories, the functions $C$ and $D$ are also supposed to depend only on $X$.

Disformal transformations induce transformations in the space of DHOST theories as follows:
\begin{equation}
    \tilde{S}[\tilde g_{\mu\nu},\phi ] = S[g_{\mu\nu}, \phi] \, ,
\end{equation}
which means that, given a theory $\tilde S$ for the metric $\tilde{g}$, one defines a new theory $S$ for the metric $g$. When the disformal transformation is invertible, the two theories are classically equivalent in vacuum only while they differ when one takes into account (generic) coupling to matter. Here, we are constructing slowly rotating solutions for  scalar-tensor theories in vacuum and then, considering the action $S$ or any disformally related action $\tilde{S}$ must lead to equivalent metrics. Let us check that this is indeed the case. 

We start with a DHOST action $S$ and we consider a slowly rotating metric, characterized by the functions $h(r)$, $f(r)$ and $\omega(r)$. We have shown  that $\omega$ satisfies the simple differential equation,
\begin{equation}
\label{eqforomega}
    \frac{\mathrm{d} \omega}{\mathrm{d}r} \; = \; \frac{k}{Q(r)} \, ,
\end{equation}
where $k$ is an integration constant and $Q(r)$ has been introduced in \eqref{Qdegen} and can be decomposed as follows,
\begin{equation}
   \label{decompQ}
        {Q} = ({Q}_2 + {Q}_3) r^4 + {Q}_B r^3 \, , \, ,
    \end{equation}
where the different components are defined by    
\begin{eqnarray}
        Q_2 & = & \sqrt{\frac{\f}{\h}}(F_2+2XA_1) \, , \\
        Q_3 & = & -\frac{1}{2}\f \sqrt{\frac{\f}{\h}}\frac{\partial \phi}{\partial r} \frac{d X}{d r}(F_{3X}+4XB_6-2B_2) \, , \\
        Q_B & = & 2 \f \sqrt{\frac{\f}{\h}} \frac{\partial \phi}{\partial r}XB_2 \, . 
    \end{eqnarray}
Notice that we have replaced $\psi'$ by the more explicit expression $\partial \phi/\partial r$ to recall that we are taking the derivative of $\phi$ at constant $t$ (since this will become important later when we change coordinates). 

Then, we perform a disformal transformation of the metric and, after a direct calculation, we can show that the metric $\tilde{g}_{\mu\nu}$ still takes the  form \eqref{hartlethornemetric}
\begin{equation}
    \mathrm{d}\tilde{s}^2 = - \tilde{\h} \, \mathrm{d}\tilde{t}^2 + \frac{\mathrm{d} \tilde{r}^2}{\tilde{\f}} + \tilde{r}^2 \, \mathrm{d}\Omega^2 - 2 \tilde r^2 \tilde{\omega}\sin^2 \theta \, \mathrm{d}\tilde{t} \, \mathrm{d} \tilde{\varphi} \, ,
\end{equation}
where the new coordinates $(\tilde{t},\tilde{r},\tilde{\varphi})$ are defined by
\begin{eqnarray}
\label{tildecoordinates}
     \mathrm{d}\tilde{t} = \mathrm{d}t - \frac{qD(r)\psi'(r)}{C(r) \h(r)-D(r)q^2}\mathrm{d}r \, , \qquad 
        \mathrm{d}\tilde{\varphi} = \mathrm{d}\varphi - \frac{q D(r)\psi'(r)\omega(r)}{C(r) \h(r)-D(r)q^2}\mathrm{d}r \, , \qquad \tilde{r}=R(r) \, ,
\end{eqnarray}
with $R(r)= \sqrt{C(r)} r$, while the new functions are given by the relations
\begin{eqnarray}
\label{relationfunctions}
    \tilde \h(\tilde r) = C \h - q^2 D \, , \qquad
    \frac{1}{\tilde \f}(\tilde r) = \frac{C}{R'^2}\left(\frac{1}{\f} + \frac{D \h \psi'^2}{C\h-q^2 D}\right) \, , \qquad \tilde\omega(\tilde r) = \omega \, ,
\end{eqnarray}
where the functions of the r.h.s. of the two identities are evaluated on $r$, viewed as a function of $\tilde{r}$.
Let us make two comments: we have used the notation $C(r)$ for $C(X(r))$ and $D(r)$ for $D(X(r))$; as $C$ is supposed to be positive, the coordinate $\tilde{r}$ is well-defined. We can formally integrate the two differential equations in \eqref{tildecoordinates} so that
\begin{eqnarray}
    \tilde{t} = t + \mu(r) \, , \qquad \tilde{\varphi} = \varphi + \nu(r) \, ,
\end{eqnarray}
where $\mu$ and $\nu$ are given in terms of integrals.
As a result, $\tilde{\omega}$ satisfies an equation similar to \eqref{eqforomega},
\begin{equation}
\label{eqforomegatilde}
    \frac{\mathrm{d} \tilde{\omega}}{\mathrm{d}\tilde r} \; = \; \frac{\tilde{k}}{\tilde Q(r)} \, ,
\end{equation}
where $\tilde{k}$ is still a constant and $\tilde{Q}$ is the same function as in \eqref{Qdegen} with the new functions $\tilde{F}_i$, $\tilde{A}_i$ and $\tilde{B}_i$ given in terms of $F_i, A_i$ and $B_i$ from disformal transformations (see the appendix of \cite{Langlois:2020xbc} for instance). 
Of course, the two equations \eqref{eqforomega} and \eqref{eqforomegatilde} must be equivalent, which leads to the consistency relation
\begin{equation}
\label{conditiondisbis}
    R' \frac{Q}{\tilde{Q}} = \frac{k}{\tilde{k}}=\text{cst} \, ,
\end{equation}
which can be verified explicitly. 

To show that this is indeed the case, we have to express $\tilde{Q}$ in terms of the coefficients of the metric $g_{\mu\nu}$ and the radial coordinate $r$. This can be done using the relation
\begin{eqnarray}
    \frac{\tilde \h}{\tilde \f} = \frac{C(C-2DX)}{R'^2}  \frac{\h}{\f} \, ,
\end{eqnarray}
and we also need to transform the derivative of $\phi$ with respect to $\tilde r$ in terms of a derivative with respect to $r$,
\begin{eqnarray}
 \frac{\partial \phi}{\partial \tilde{r}} = \frac{\partial}{\partial \tilde{r}} \left[ q\tilde t + q \mu(r) + \psi(r) \right] =  \frac{1}{R'}(q \mu' + \psi')=C \frac{\h}{\tilde{\h}} \frac{\psi'}{R'} \, . 
\end{eqnarray}
After a direct calculation, we can see that $\tilde{Q}$ takes the same form as \eqref{decompQ},
\begin{eqnarray}
    \tilde{Q} = (\tilde{Q}_2 + \tilde{Q}_3) r^4 + \tilde{Q}_B r^3 \, ,
\end{eqnarray}
where the three components are now given by  
\begin{eqnarray}
    \tilde{Q}_2 & = & R' \sqrt{\frac{\f}{\h}} \frac{C^{3/2}}{(C-2DX)^{1/2}} (\tilde{F}_2 +2 \tilde{X} \tilde{A}_1) \, , \\
    \tilde{Q}_3 & = & -\frac{R'}{2} \f \sqrt{\frac{\f}{\h}}  \psi'X' \frac{ C^{3/2}}{(C-2DX)^{3/2}}  \frac{\partial \tilde X}{\partial X}\left[ \tilde{F}_{3\tilde X} + 4 \tilde{X} \tilde{B}_6 -2 (1+ \frac{\tilde{X} C_{\tilde X}}{C} ) \tilde{B}_2 \right]\, , \\
   \tilde{Q}_B & = & 2R' \sqrt{\frac{\f}{\h}}  \f \psi' \frac{C^{3/2}}{(C-2DX)^{3/2}} \tilde{X} \tilde{B}_2 \, ,
\end{eqnarray}
with $\tilde{X}= X/(C-2XD)$. As a consequence, we have to show that
\begin{eqnarray}
\label{ratio}
    R' \frac{Q_2}{\tilde{Q_2}} = R' \frac{Q_3}{\tilde{Q}_3} = R' \frac{Q_B}{\tilde{Q}_B} = \text{cst} \, .
\end{eqnarray}
In the case of quadratic theories, only $Q_2$ is non trivial, and verifying  the relation between $Q_2$ and $\tilde{Q}_2$ is easy and can be done from a direct calculation using disformal transformations formulae \cite{Langlois:2020xbc},
\begin{eqnarray}
    F_2=(C(C-2DX))^{1/2} \tilde{F}_2 \, , \qquad
    A_1 = \frac{DC^{1/2}}{(C-2DX)^{1/2}} \tilde{F}_2 + \frac{C^{3/2}}{(C-2DX)^{3/2}} \tilde{A}_1 \, ,
\end{eqnarray}
which leads immediately to the relation
\begin{eqnarray}
    F_2 + 2X A_1 = \frac{C^{3/2}}{(C-2DX)^{1/2}} (\tilde{F}_2 + 2 \tilde{X} \tilde{A}_1) \, .
\end{eqnarray}

Furthermore, we see that the constant entering in \eqref{ratio} is equal to 1, which  implies that the two integration constants $k$ and $\tilde{k}$ \eqref{eqforomegatilde} are identical. 

This result can be extended to cubic theories. In that case, we assume that $\tilde S$ is the Horndeski action so that we can use the disformal transformations formulae given in section V of \cite{BenAchour:2016fzp}.
In that case, $\tilde{F}_3$ is an arbitrary function, and
\begin{eqnarray}
    \tilde{B}_2 =  \frac{1}{2}\tilde{F}_{3\tilde X} \, , \qquad \tilde{B}_6=0 \, ,
\end{eqnarray}
so that, after a few calculations, we have,
\begin{eqnarray}
    \tilde Q_3= - \frac{R'}{2} \f \sqrt{\frac{\f}{\h}} \psi' X' \frac{X C_X C^{1/2}}{(C-2D X)^{5/2}} \frac{\partial X}{\partial \tilde{X}} \tilde{F}_{3X} \, .
\end{eqnarray}
The functions entering in the disformed action $S$ are obtained from disformal transformations relations which imply here,
\begin{eqnarray}
    &&F_{3X} = \frac{C^{1/2}}{(C-2DX)^{1/2}} \tilde{F}_{3X} \, , \qquad B_2=\frac{C^{3/2} \, \tilde{F}_{3X}}{2(C -2X D)^{1/2} (C - X C_X +2X^2 D_X)} \, , \nonumber \\
    && B_6 = -\frac{C^{1/2} X D_X \, \tilde{F}_{3X}}{2(C-2X D)^{1/2} (C-X C_X + 2X^2 D_X)} \, .
\end{eqnarray}
Thus, we can easily compute $Q_3$ and $Q_B$, and check that the remaining two relations \eqref{ratio} are satisfied with the constant being 1 as well. 

There is an immediate consequence of this consistency relations. Given a DHOST action $S$ which admits a slowly rotating solution associated with the function $\f$, $\h$ and $\omega$, then the disformally related action $\tilde S$ admits also a slowly rotating solution associated to $\tilde \f$, $\tilde \h$ and $\tilde \omega$ related to the original ones by \eqref{relationfunctions}. In particular, the frame dragging functions are related by
\begin{equation}
    \omega(r)=\tilde{\omega}\left( R(r)\right) \, , \quad \text{with} \quad R(r) = r\sqrt{C(r)} \, .
\end{equation}
Expressing $\tilde{\omega}$ in terms of $\omega$ can be done only if the coordinate transformation $\tilde r= \sqrt{C(r)}r$ is invertible, or equivalently $\tilde r$ is a monotonous function of $r$. 

\section{More details on the equation for an angular dependent frame dragging function}
\label{wrthappendix}
\hspace{\parindent} In this appendix, we give more details on the equation \eqref{eqwrth} with \eqref{Qtheta} for a $\theta$-dependent $\omega$. Following the method described in section \ref{shiftsymmetryargument}, the equation \eqref{conservationequation} now becomes 
    \begin{equation}
    \label{dSdw2}
        \frac{\delta S}{\delta\omega}=-\frac{\partial}{\partial r}\left(\frac{\delta S}{\delta \partial_r\omega}\right)-\frac{\partial}{\partial\theta}\left(\frac{\delta S}{\delta \partial_\theta\omega}\right) \, ,
    \end{equation}
    where the two derivatives  can be computed directly from the DHOST Lagrangian, 
    \begin{eqnarray}
    \label{dSdwr}
        \frac{\delta S}{\delta \partial_r\omega} &=& Q(r)\sin^3(\theta)\frac{\partial\omega}{\partial r}+\frac{1}{2}\frac{\partial}{\partial\theta}\left(\sqrt{\frac{f}{h}}r^2\psi'F_3\sin^3(\theta)\frac{\partial\omega}{\partial\theta}\right) \, , \\
         \frac{\delta S}{\delta \partial_\theta\omega} &=& \frac{1}{\sqrt{hf}}\bigg[\left(F_2+\frac{q^2}{h}A_1\right)r^2+f\psi'\frac{q^2}{h}(3B_3+2B_2)r+\frac{1}{2}f\psi'X'\left(F_{3X}-2\frac{q^2}{h}B_6\right)r^2 \nonumber \\
         && -\frac{q^2}{h}B_2\frac{(hX)'}{h\psi'}r^2\bigg]\sin^3(\theta)\frac{\partial\omega}{\partial\theta}-\frac{1}{2}\frac{\partial}{\partial r}\left(\sqrt{\frac{f}{h}}r^2\psi'F_3\sin^3(\theta)\frac{\partial\omega}{\partial\theta}\right) \, .\label{dSdwth}
    \end{eqnarray} 
    Notice that $Q(r)$ is the same function as the one introduced in \eqref{Qnondegen}. 
    
    Remarkably, the last term in \eqref{dSdwr} cancels with the last term in \eqref{dSdwth} when combined in \eqref{dSdw2}, yielding an equation of the form of \eqref{eqwrth}. Finally, simplifying the expression of \eqref{dSdwth} using the degeneracy condition $3B_3+2B_2=0$ gives  \eqref{Qtheta}.

    \bibliographystyle{JHEP}
    \bibliography{biblio.bib}
\end{document}